# Van der Pauw method on a sample with an isolated hole

Krzysztof Szymański, Jan L. Cieśliński and Kamil Łapiński

Uniwersytet w Białymstoku, Wydział Fizyki,
ul. Lipowa 41, 15-424 Białystok, Poland

email: kszym@alpha.uwb.edu.pl, tel/fax no : (48)857457221/23

Abstract

Explicit results of the van der Pauw method for a sample containing an isolated hole are presented together with experimental confirmation. Results of measurements and numerical analysis strongly suggest that four probe resistivities obey inequality similar in the form to the famous van der Pauw equation. The inequality seems to be valid for any sample with an isolated hole and contacts located on the same edge, however rigorous proof is not given. The inequality can be used for experimental detection of the sample quality.

**1. Introduction**

Van der Pauw has shown that appropriately defined resistivities for any arrangement of four point contacts on the edge of a flat homogeneous sample obey a functional relation (an implicit function of two resistivities, see eq. (1) in [1]). It can be proved in a simple and elegant way using conformal mappings of complex functions. In a recent paper [3] it has been pointed out that choosing pairs of the resistivities in different ways one obtains additional two equations, having essential advantages in numerical computations. The van der Pauw relation has important practical applications since it allows measurement of specific resistivity and Hall coefficient for samples of arbitrary shape [1, 2]. The van der Pauw geometry has been successfully used in transport measurements of massless electrons [4], quantum Hall effect [5] in graphene, transport of carriers in quantum well [6], extraordinary magnetoresistance in graphene annulus with metallic disc [7] and high $T_c$ superconductors [8]. The original van der Pauw idea was recently applied also for heat transport measurements [9,10]. Using mapping theory, the results of van der Pauw were generalized to an anisotropic medium [11]. A rectangular anisotropic samples were investigated theoretically and experimentally in [12]. In ideal van der Pauw measurements the sample is assumed to be homogeneous, has to have a constant and small thickness and infinitesimal contacts at the edge. Problem of finite contact size and their out-of-edge location was first analyzed in [2]. Influence of non-ideal conditions is still debated. Effect of contacts location, sample inhomogeneities and holes were considered further in [13-17]. In this paper the van der Pauw method is applied to a sample with an isolated hole.

## 2. Four probe resistivities for a sample with an isolated hole and for a cylinder

Let the current $j_{PQ}$ flows from $P$ to $Q$ and the potential is measured between point contacts $R$ and $S$, located on the same edge of a thin sample (holes are admitted). We define resistivities $R_{PQ,RS}=(V_S-V_R)/j_{PQ}$ in the same way as in [1]. We also denote

$$v \equiv \exp(-R_{PQ,RS}/\lambda) + \exp(-R_{QR,SP}/\lambda), \qquad (1)$$

where $\lambda=\rho/(\pi d)$ and $\rho$ is specific resistivity of the sample with thickness $d$. The van der Pauw relation (eq. (1) in [1]) states $v=1$ and is valid only for samples without holes. We postulate a hypothesis that in the general case (samples with holes) $v$ fulfills the following inequality:

$$v \leq 1. \qquad (2)$$

We are going to support this conjecture by experimental data (see the next section) and by detailed analysis of the one-hole case.

The disc with a hole can be conformally mapped to the cylinder of finite height, for more details see the last section. Then, the cylinder can be identified with a periodic lattice. We consider two dimensional lattice of simple rectangular type [18] and a primitive cell composed of the current source and the sink, see (Fig. 1). Unit cell dimensions are $p$ and $2H$. The electric potential $V_0$ at point $(x,y)$ from the single current source located at $(-a/2,0)$ and the sink located at $(+a/2,0)$ is given by:

$$V_0(x,y) = \frac{j\rho}{\pi d} \ln\left(\frac{(2x-a)^2 + 4y^2}{(2x+a)^2 + 4y^2}\right)^{1/2}. \qquad (3)$$

The potential at point $(x,y)$ resulting from the lattice is equal to the lattice sum:

$$V(x,y) = \sum_{\substack{n_x=-\infty \\ n_y=-\infty}}^{\infty} V_0(x - pn_x, y - 2Hn_y). \qquad (4)$$

In order to determine the potential we applied the method of images [19]. Vertical lines $x=-p/2$, $x=+p/2$ are equipotential lines. Along horizontal lines given by $y=0$ and $y=H$ the current flows only in $x$-direction. Therefore the unit cell can be wound into a cylinder with a finite height $H$ and circumference $p$ (Fig. 1a). Evaluating $V(x,y)$ at $y=0$ we get the potential produced by $j_{PQ}$ ($x$ is related to positions of contacts on the edge). Next, we introduce dimensionless variables $\alpha=2\pi a/p$, $\beta$, $\gamma$ and $h=4\pi H/p$, see Fig. 1a. Finally, after straightforward calculations, we obtain four probe resistivities in an explicit form:

$$R_{PQ,RS} = \lambda \ln\left|\frac{G(\alpha+\beta)G(\alpha+\gamma)}{G(\beta)G(\gamma)}\right|,$$

$$R_{QR,SP} = \lambda \ln\left|\frac{G(\alpha+\beta)G(\alpha+\gamma)}{G(\alpha)G(\delta)}\right|, \qquad (5)$$

$$R_{RP,QS} = \lambda \ln\left|\frac{G(\alpha)G(\delta)}{G(\beta)G(\gamma)}\right|,$$

where

$$G(\phi) = \sin\frac{\phi}{2} \prod_{n=1}^{\infty}\left(1 - \frac{\cos\phi}{\cosh hn}\right), \qquad (6)$$

and $\delta=2\pi-\alpha-\beta-\gamma$, $h=4\pi H/p$. We assume $\alpha$, $\beta$, $\gamma$ positive and $\alpha+\beta+\gamma\leq 2\pi$, see Fig. 1a. Resistivities given by Eq. (5) satisfy the reciprocity theorem, in particular relations (5-8) given in [1]: $R_{PQ,RS}=R_{RS,PQ}$, $R_{QR,SP}=R_{SP,QR}$, $R_{PR,QS}=R_{QS,PR}$, $R_{PQ,RS}+R_{QR,SP}+R_{PR,QS}=0$.

In the case of a finite cylinder the quantity $v$, defined by formula (1), can be expressed in terms of one function $G(\phi,h)$, see eq. (6). Then formulae (1), (2) assume the form:

$$v = \frac{G(\alpha)G(\delta) + G(\beta)G(\gamma)}{G(\alpha + \beta)G(\alpha + \gamma)} \leq 1. \tag{7}$$

However, we could not find complete mathematical proof yet. Numerical checks show that this inequality is satisfied for any $\alpha$, $\beta$, $\gamma$ such that $\alpha+\beta+\gamma \leq 2\pi$. An example of two dimensional plot $v(\alpha,\beta,\gamma)$ with additional constraints constrains $\beta=\gamma$ and arbitrarily chosen $h=1.1$ to decrease the number of variables, is shown in Fig. 2 to support this statement. In the limit of large $h$ (the case of long cylinder or a flat sample with a very small hole) we have:

$$v = 1 - 16 \sin\frac{\alpha}{2} \sin\frac{\beta}{2} \sin\frac{\gamma}{2} \sin\frac{\delta}{2} e^{-h} + O(e^{-2h}). \tag{8}$$

Hence, $v \to 1$ for $h \to \infty$. Moreover, eq. (8) implies that there exists $h_1$ such $v<1$ for $h>h_1$. Thus we proved inequality (2) at least in the case of sufficiently large $h$.

## 3. Experimental results

The first experiment was performed on a flat sample made of 0.1mm thick brass without isolated holes. Because of macroscopic dimensions (see left-hand part of the inset in Fig. 3), it was quite easy to arrange contacts at any positions on the edge. To demonstrate precision of the experiment correlation between two resistivities were measured for many different contacts arrangements. All experimental points in Fig. 3 (boxes) lie on the line $v=1$ which was drawn without any adjustable parameters. The specific resistivity was measured in separate experiment on a long stripe made of the same brass sheet.

In the second experiment a hole was made in the brass sheet sample (see right hand part of the inset in Fig. 3). All contacts were arranged on the outer edge. Correlation between resistivities is shown in Fig. 3. Usually, for arbitrary arrangements of contacts, the experimental points (diamonds) lie above the curve $v=1$, consistently with inequality (2). However, when the contacts were placed close to each other, the points in Fig. 3 are near to this curve showing that the obtained correlation is close to the van der Pauw relation. This is actually quite intuitive. If the contacts are close to each other the hole is essentially far away from any flowing current and does not affect the measurement.

In the third series of measurements we checked predictions of eq. (5). A cylindrical sample with diameter 178 mm, height $H=30$ mm and wall thickness $d=1$ mm was prepared from commercially available stainless steel tube. Arranging all contacts on one edge in symmetric manner, e.g. $\alpha=\beta=\gamma$, we measured relevant resistivities. Results are shown in Fig. 4. Part of the tube was cut into long and narrow stripe for independent measurement of the ratio of the specific resistivity and the thickness. For the obtained value $\rho/d=0.711$ m$\Omega$, predictions of eq. (5) are shown by solid lines with excellent agreement between theory and experiment in Fig. 4.

The experimental results shown in Fig. 3 and 4 were obtained on student laboratory setup. Measurements were performed at ambient conditions without temperature stabilization. Because of the reciprocity theorem, linear combination with coefficients $\pm 1$ of three resistivities (5) should vanish. We always measured all three resistivities and their appropriate linear combination served as a measure of the experimental precision. In actual experiments the relative precision of resistivities measurements (shown in Figs. 3 and 4) was not worse than 0.1%.

## 4. Discussion and conclusions

The finite cylinder was constructed from a rectangular cell (see Fig. 1b) which can be considered as a subset $|\text{Re}(z)|<p/2$ and $0<\text{Im}(z)<H$ of the complex plane $z=x+iy$. One may perform conformal mapping $z\rightarrow\exp(2\pi i z/p)$ of the rectangle into the annulus, centered at 0, of outer radius 1 and inner radius $\exp(-2\pi H/p)$ (i.e., the unit disc with a central hole). Therefore, if inequality (2) is true for a cylinder, it is also true for the annulus and all other shapes obtained by conformal mappings. We note however, that in contrast to samples without holes (all simply connected regions are conformally equivalent), two samples with single holes in general are not conformally equivalent. Even two discs with concentric single holes cannot be mapped conformally into each other unless the holes are of the same size, see [20], section 14.21. Nevertheless any disc with a circular hole can be transformed by a conformal mapping into an annulus and then corresponding resistivities are given by formulae (5). Indeed, it is easy to check that unit circle with circular hole of radius $r$ and eccentricity $e$ can be transformed by conformal mapping $f$:

$$f(z) = \frac{z+\xi}{z\xi+1}, \tag{11}$$

$$\xi = \frac{1}{2e}\left(r^2 - e^2 - 1 + \sqrt{((1-e)^2 - r^2)((1+e)^2 - r^2)}\right), \tag{12}$$

into an annulus with the outer radius equal to 1 and inner radius equal to $f(e+r)$. By eccentricity we understand a distance between origin and the center of the hole.

Let us consider a dependence of $\nu$ on the location of contacts. The obvious case of the contacts approaching to each other, results in the upper limit of $\nu$ equal to 1. In the case of a cylinder (or a disc with a central hole) the minimum value of $\nu$ is realized for the symmetric arrangement of contacts ($\alpha=\beta=\gamma=\pi/2$), compare (8). Then

$$\nu_{\min} = G^{-2}(\pi) = \frac{1}{\prod_{n=1}^{\infty}(1+\frac{1}{\cosh nh})}. \tag{13}$$

In the case of small hole or long cylinder (which is equivalent to taking a large $h$) the minimum value is approximated by

$$\nu_{\min} = 1 - 4e^{-h} + O(e^{-2h}). \tag{14}$$

This minimum value can be interpreted as a measure of the disturbance caused by the hole. If the exact location of the hole is not known then, in order to estimate $\nu_{\min}$, one has to perform many measurements for different arrangements of contacts.

To our best knowledge formulae (5) are the first exact and explicit result related to the van der Pauw method for a sample with an isolated hole. It is worthwhile to point out that measuring van der Pauw resistivities and using inequality (2) one may detect the presence of a hole in the sample (provided that the sheet resistance of the sample is known and isotropic). Thus inequality (2) can be used for detection of inhomogeneities in the sample or for determination of the sample quality.

**Acknowledgements**
The authors thank to students G. Kulesza and Ł. Wieleszczyk for collaboration during measurements on the students laboratory setup. Thanks are also due to anonymous reviewers for valuable comments.

**References**
[1] L. J. van der Pauw: A method of measuring specific resistivity and Hall effect of discs of arbitrary shape, Philips Research Reports **13,** 1 (1958).
[2] L. J. van der Pauw: A method of measuring the resistivity and Hall coefficient on lamellae of arbitrary shape, Philips Technical Review **20,** 220 (1958).
[3] J. L. Cieśliński: Modified van der Pauw method based on formulas solvable by the Banach fixed point method, Thin Solid Films, **522**, 314 (2012).
[4] J. R. Williams and C. M. Marcus, Snake States along Graphene p-n Junctions, Phys. Rev. Lett. **107,** 046602 (2011).
[5] Z. Jiang, Y. Zhang, Y.-W. Tan, H.L. Stormer and P. Kim, Quantum Hall effect in graphene, Solid State Communications **143,** 14 (2007).
[6] S. Dasgupta, S. Birner, C. Knaak, M. Bichler, A. Fontcuberta i Morral, G. Abstreiter and M. Grayson, Single-valley high-mobility (110) AlAs quantum wells with anisotropic mass, Appl. Phys. Lett. **93,** 132102 (2008).
[7] J. Lu, H. Zhang, W. Shi, Z. Wang, Y.Zheng, T. Zhang, N. Wang, Z. Tang and P. Sheng, Graphene Magnetoresistance Device in van der Pauw Geometry, Nano Letters **11**, 2973 (2011).
[8] W. Preis, M. Holzinger and W. Sitte, Application of the van der Pauw Method to Conductivity Relaxation Experiments on $YBa_2Cu_3O_{6+\delta}$, Monatshefte fur Chemie **132**, 499 (2001).
[9] J. de Boor and V. Schmidt, Complete characterization of thermoelectric materials by a combined van der Pauw approach, Advanced Materials **22**, 4303 (2010).
[10] J. de Boor and V. Schmidt, Efficient thermoelectric van der Pauw measurements, Appl. Phys. Lett. **99,** 022102 (2011).
[11] H. Shibata and R. Terakado, A potential problem for point contacts on a two dimensional anisotropic medium with an arbitrary resistivity tensor, J. Appl. Phys. **66**, 4603 (1989).
[12] O. Bierwagen, R. Pomraenke, S. Eilers, and W. T. Masselink, Mobility and carrier density in materials with anisotropic conductivity revealed by van der Pauw measurements, Phys. Rev. B **70**, 165307 (2004).
[13] O. Bierwagen, T. Ive, C.G. Van de Walle, and J. S. Speck, Causes of incorrect carrier-type identification in van der Pauw–Hall measurements, Appl. Phys. Lett.7 93, (2008) 242108.
[14] D. W. Koon and C. J. Knickerbocker, What do you measure when you measure resistivity?, Rev. Sci. Instrum. **63**, 207 (1992).
[15] T.Ohgaki, N.Ohashi, S.Sugimura, H.Ryoken, I.Sakaguchi, Y.Adachi and H.Haneda, Positive Hall coefficients obtained from contact misplacement on evident n-type ZnO films and crystals, J. Mater. Res. **23**, 2293 (2008).

[16] B. Wu, X. Huang, Y. Han, C. Gao, G. Peng, C. Liu, Y. Wang, X. Cui, and G. Zou, Finite element analysis of the effect of electrodes placement on accurate resistivity measurement in a diamond anvil cell with van der Pauw technique, J. Appl. Phys. **107,** 104903 (2010).
[17] J. Nahlık, I. Kasparkova, P. Fitl: Study of quantitative influence of sample defects on measurements of resistivity of thin films using van der Pauw method, Measurement **44**, 1968 (2011).
[18] C. Kittel: Introduction to Solid State Physics, chapter 1, 8th Edition, Wiley, 2005.
[19] R. Feynman, R. Leighton, and M. Sands: The Feynman Lectures on Physics, vol. 2, Addison Wesley Longman, 1970.
[20] W. Rudin: Real and Complex Analysis, McGraw-Hill, 1974.

# Figure captions

Fig. 1. a) Cylinder with four contacts on the same edge, b) rectangular lattice of current sources and sinks corresponding to the cylinder (see section 2).

Fig. 2. 3-dimensional plot (left) and contours (right) of $1-v(\alpha,\beta,\gamma)$ with additional constraints $\beta=\gamma$ and $h=1.1$. Vertical separation between contours is 0.1.

Fig. 3. Correlation between $R_{PQRS}$ and $R_{QRSP}$ measured on a flat sample (squares) with different configurations of the contacts on the edge. The line shows relation $v=1$. Diamonds show correlation between $R_{PQRS}$ and $R_{QRSP}$ measured on a sample with the same material and with a hole. Inset: shapes of the samples used in the experiments.

Fig. 4. Four-point resistivities $R_{PQRS}$ (red), $R_{QRSP}$ (green) and $R_{PRQS}$ (blue), measured with contact located on one edge of the cylinder with equidistant contacts ($\alpha=\beta=\gamma$), see eq. (2). Dots are results of measurements while solid lines represent eq. (2). See Fig. 2 for a scheme of the experiment.

# Figures

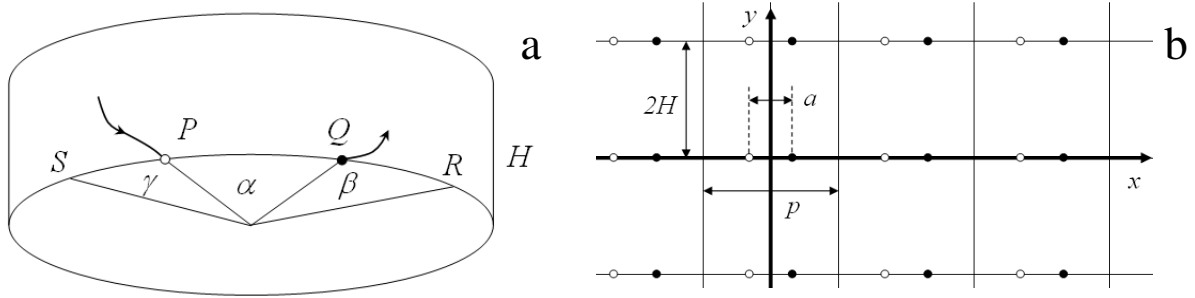

Fig. 1. a) Cylinder with four contacts on the same edge, b) rectangular lattice of current sources and sinks corresponding to the cylinder (see section 2).

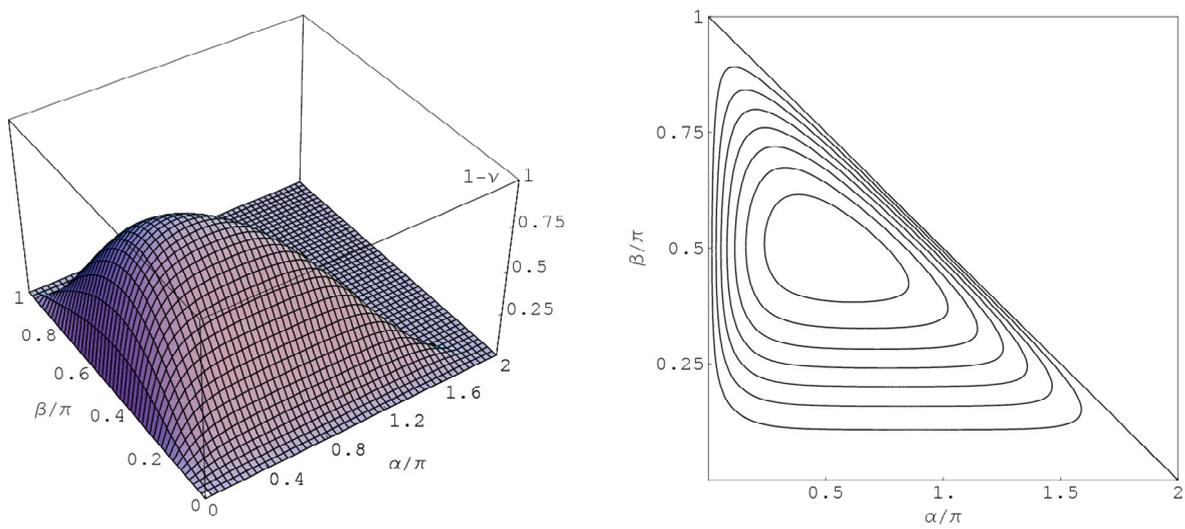

Fig. 2. 3-dimensional plot (left) and contours (right) of $1-v(\alpha,\beta,\gamma)$ with additional constraints $\beta=\gamma$ and $h=1.1$. Vertical separation between contours is 0.1.

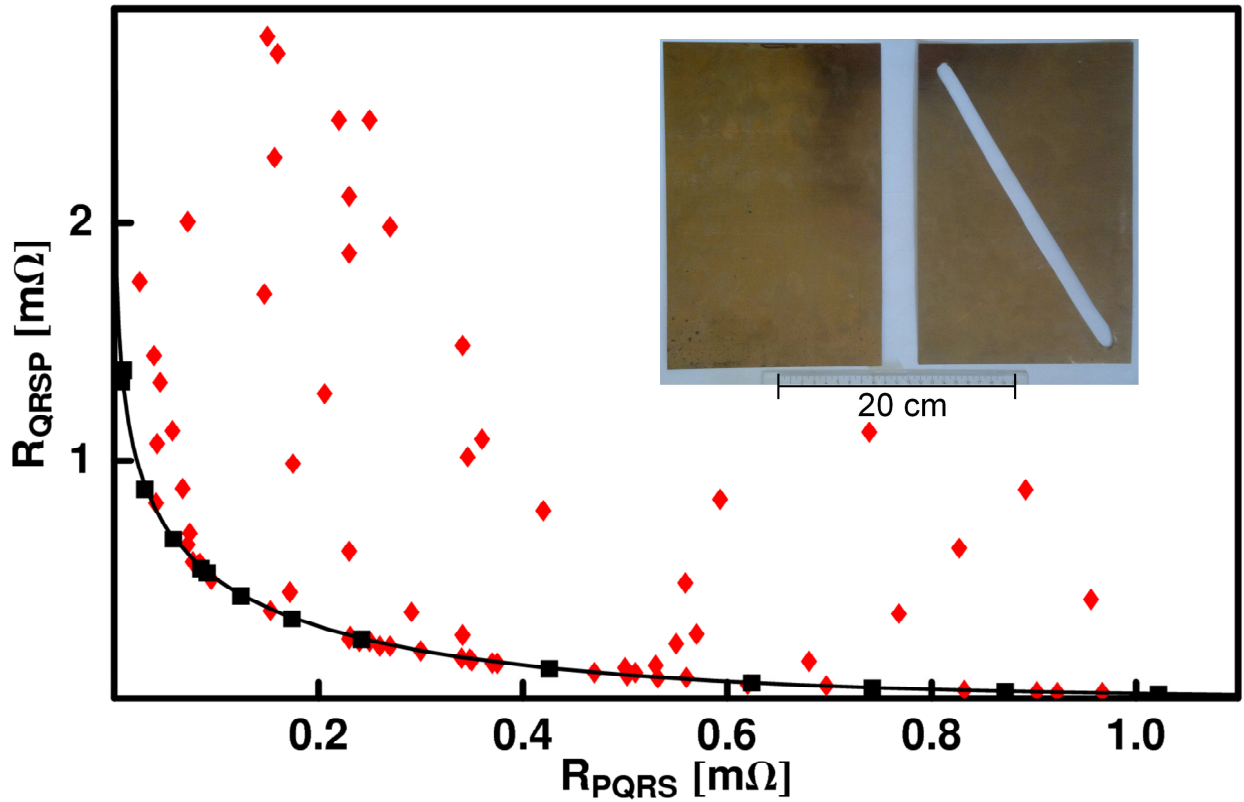

Fig. 3. Correlation between $R_{PQRS}$ and $R_{QRSP}$ measured on a flat sample (squares) with different configurations of the contacts on the edge. The line shows relation $\nu=1$. Diamonds show correlation between $R_{PQRS}$ and $R_{QRSP}$ measured on a sample with the same material and with a hole. Inset: shapes of the samples used in the experiments.

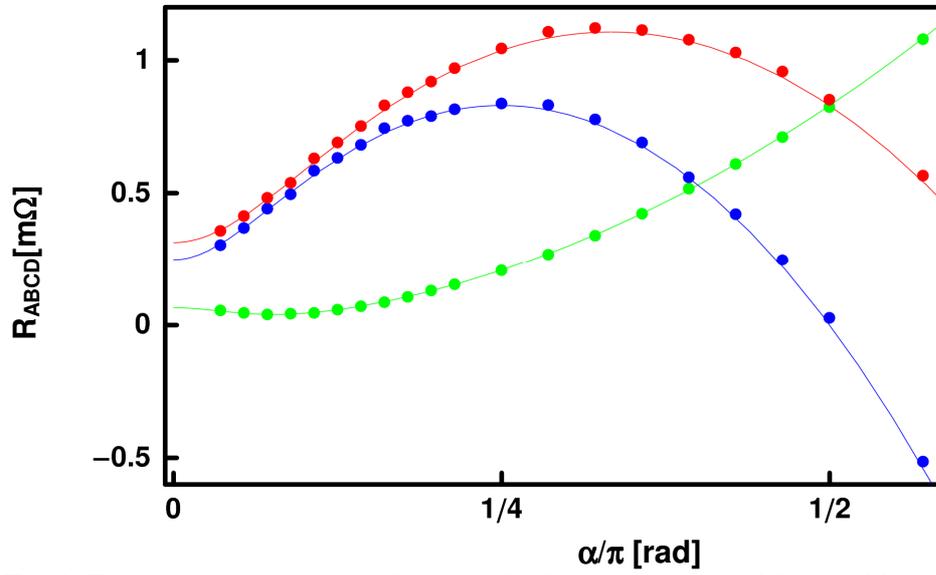

Fig 4. Four-point resistivities $R_{PQRS}$ (red), $R_{QRSP}$ (green) and $R_{PRQS}$ (blue), measured with contact located on one edge of the cylinder with equidistant contacts ($\alpha=\beta=\gamma$), see eq. (2). Dots are results of measurements while solid lines represent eq. (2). See Fig. 1 for a scheme of the experiment.